# Static Scaling Behavior of High-Molecular-Weight Polymers in Dilute Solution: A Reexamination

Alan D. Sokal
*Department of Physics*
*New York University*
*4 Washington Place*
*New York, NY 10003 USA*
SOKAL@NYU.EDU

April 27, 1993

### Abstract

Previous theories of dilute polymer solutions have failed to distinguish clearly between two very different ways of taking the long-chain limit: (I) $N \to \infty$ at fixed temperature $T$, and (II) $N \to \infty$, $T \to T_\theta$ with $x \equiv N^\phi(T - T_\theta)$ fixed. I argue that the modern two-parameter theory (continuum Edwards model) applies to case II — not case I — and in fact gives exactly the crossover scaling functions for $x \geq 0$ modulo two nonuniversal scale factors. A Wilson-type renormalization group clarifies the connection between crossover scaling functions and continuum field theories.

For several decades, most work on the behavior of long-chain polymer molecules in dilute solution [1, 2, 3, 4, 5, 6] has been based on the so-called "two-parameter theory" in one or another of its variants: traditional (Flory-type) [7], pseudo-traditional (modified Flory-type) [8, 9] or modern (continuous-chain-type) [4, 5]. In this Letter I shall argue that *all such theories are wrong*. They are wrong not merely because they make incorrect predictions, but for a more fundamental reason: they purport to make universal predictions for quantities that are not in fact universal. (Very similar ideas have been expressed in an important recent paper of Nickel [10].) However, I shall also argue that the modern two-parameter theory has a valid *reinterpretation* as a theory of the universal crossover scaling behavior in an infinitesimal region just above the theta temperature.

The universal properties of polymer molecules manifest themselves in the long-chain limit $N \to \infty$, where $N$ is the number of monomers in the chain. However, it is crucial to distinguish two *very different* ways of taking this limit:

(I) $N \to \infty$ at fixed temperature $T$, where either (a) $T > T_\theta$, (b) $T = T_\theta$, or (c) $T < T_\theta$.

(II) $N \to \infty$, $T \to T_\theta$ with $x \equiv N^\phi (T - T_\theta)$ fixed, where $\phi$ is a suitable *crossover exponent*.

The good-solvent (GS) regime corresponds to case Ia. Here standard renormalization-group (RG) arguments predict [10] that the mean-square end-to-end distance $\langle R^2 \rangle$, the mean-square radius of gyration $\langle S^2 \rangle$ and the second virial coefficient $A_2^{(mol)} \equiv (N^2 M_{monomer}^2 / N_{Avogadro}) A_2$ have the asymptotic behavior

$$\langle R^2 \rangle = B_R N^{2\nu_{GS}} (1 + b_R^{(1)} N^{-\Delta_{1,GS}} + \ldots) \quad (1)$$

$$\langle S^2 \rangle = B_S N^{2\nu_{GS}} (1 + b_S^{(1)} N^{-\Delta_{1,GS}} + \ldots) \quad (2)$$

$$A_2^{(mol)} = B_A N^{d\nu_{GS}} (1 + b_A^{(1)} N^{-\Delta_{1,GS}} + \ldots) \quad (3)$$

as $N \to \infty$, where $d$ is the spatial dimension. The critical exponents $\nu_{GS}$ and $\Delta_{1,GS}$ are universal. The amplitudes $B_R, B_S, B_A, b_R^{(1)}, b_S^{(1)}, b_A^{(1)}$ are nonuniversal; in fact, even the *signs* of the correction-to-scaling amplitudes $b_R^{(1)}, b_S^{(1)}, b_A^{(1)}$ [and their various combinations such as $b_\Psi^{(1)} \equiv b_A^{(1)} - (d/2) b_S^{(1)}$] are nonuniversal. However, the RG theory also predicts that the dimensionless amplitude *ratios* $B_S/B_R$, $B_A/B_R^{d/2}$, $b_S^{(1)}/b_R^{(1)}$ and $b_A^{(1)}/b_R^{(1)}$ are universal [10, 11].

Thus, in the continuum Edwards model [12] ($\simeq$ continuum $\varphi^4$ field theory with $n=0$ components), the effective exponents $\nu_{eff,R} \equiv \frac{1}{2} d \log\langle R^2 \rangle / d \log N$ and $\nu_{eff,S} \equiv \frac{1}{2} d \log\langle S^2 \rangle / d \log N$ and the interpenetration ratio $\Psi \equiv 2(d/12\pi)^{d/2} A_2^{(mol)} / \langle S^2 \rangle^{d/2}$ all approach their asymptotic values *from below* [4, 5, 13, 14]: that is, $b_R^{(1)}, b_S^{(1)} > 0$ and $b_\Psi^{(1)} < 0$. On the other hand, recent high-precision Monte Carlo studies of lattice self-avoiding walks [10, 15] show clearly that these quantities approach their asymptotic values *from above*; and the same occurs in the bead-rod model with sufficiently large bead diameter [16]. Indeed, this latter behavior is almost obvious qualitatively: short self-avoiding walks behave roughly like hard spheres; only at larger $N$ does one see the softer excluded volume (smaller $\Psi$) characteristic of a fractal object. In any case, all these models are in excellent agreement for the leading *universal* quantities $\nu_{GS}$, $B_S/B_R$ and $\Psi^* \equiv 2(d/12\pi)^{d/2} B_A/B_S^{d/2}$, and they are in rough agreement for the universal correction-to-scaling quantities $\Delta_{1,GS}$, $b_S^{(1)}/b_R^{(1)}$ and $b_A^{(1)}/b_R^{(1)}$.

It is thus misguided to analyze the experimental data in the good-solvent regime by attempting to match the real polymer molecules to the continuum Edwards model via the correspondence $z_{Edwards} = aN^{1/2}$ (where $a$ is an empirically determined scale factor depending on the polymer, solvent and temperature) [17]: the continuum Edwards model can predict *only* the universal quantities. Indeed, there is evidence [18] that real polymers in a sufficiently good solvent behave like self-avoiding walks, i.e. they approach $\Psi^*$ from above; in this case they *cannot* be matched to *any* value of $z_{Edwards}$. This behavior has heretofore been considered paradoxical; in fact, it is quite natural [19].

These points have been made previously by Nickel [10]. Similar comments have been made with regard to liquid-gas critical points by Liu and Fisher [20].

A very different limiting behavior is obtained (in dimension $d < 4$) if we take simultaneously $N \to \infty$ and $T \to T_\theta$ such that $x \equiv N^\phi(T - T_\theta)$ remains fixed. For suitably chosen exponents $\phi$ and $\nu_\theta$, the following limits are expected to exist:

$$f_R(x) \equiv \lim_{\substack{N \to \infty \\ T \to T_\theta \\ x \equiv N^\phi(T-T_\theta) \text{ fixed}}} \frac{\langle R^2 \rangle_{N,T}}{N^{2\nu_\theta}} \tag{4}$$

$$f_S(x) \equiv \lim_{\substack{N \to \infty \\ T \to T_\theta \\ x \equiv N^\phi(T-T_\theta) \text{ fixed}}} \frac{\langle S^2 \rangle_{N,T}}{N^{2\nu_\theta}} \tag{5}$$

$$f_A(x) \equiv \lim_{\substack{N \to \infty \\ T \to T_\theta \\ x \equiv N^\phi(T-T_\theta) \text{ fixed}}} \frac{A_2^{(mol)}(N,T)}{N^{d\nu_\theta}} \tag{6}$$

The exponents $\phi$ and $\nu_\theta$ are universal, and the *crossover scaling functions* $f_R$, $f_S$ and $f_A$ are universal modulo a rescaling of abscissa and ordinate. The exponents are believed to take the values

$$\phi = \begin{cases} 2 - \frac{d}{2} & \text{for } 3 < d < 4 \\ \frac{1}{2} \times \log^{3/22} & \text{for } d = 3 \text{ [21]} \\ \frac{1}{2} + \frac{3\epsilon}{22} + \ldots & \text{for } d = 3 - \epsilon \text{ [21, 22]} \\ \frac{3}{7} & \text{for } d = 2 \text{ [23]} \end{cases} \tag{7}$$

$$\nu_\theta = \begin{cases} \frac{1}{2} & \text{for } d \geq 3 \\ \frac{1}{2} + \frac{2\epsilon^2}{363} + \ldots & \text{for } d = 3 - \epsilon \ [21, 22] \\ \frac{4}{7} & \text{for } d = 2 \ [23] \end{cases} \qquad (8)$$

The functions $f_R$ and $f_S$ (and $f_\Psi \sim f_A/f_S^{d/2}$ at least for $x \geq 0$) are monotonically increasing functions of their argument $x \equiv N^\phi(T - T_\theta)$, with the asymptotic behavior

$$f_R(x), f_S(x) \sim \begin{cases} x^{2(\nu_{GS} - \nu_\theta)/\phi} & \text{as } x \to +\infty \\ (-x)^{2(\nu_{coll} - \nu_\theta)/\phi} & \text{as } x \to -\infty \end{cases} \qquad (9)$$

$$f_A(x) \sim \begin{cases} x^{d(\nu_{GS} - \nu_\theta)/\phi} \ (>0) & \text{as } x \to +\infty \\ \text{unknown} \ (<0) & \text{as } x \to -\infty \end{cases} \qquad (10)$$

where $\nu_{coll} = 1/d$. In fact, I shall argue below that, for $3 \leq d < 4$, the functions $f_R(x)$, $f_S(x)$ and $f_A(x)$ for $x \geq 0$ are given precisely by the continuum Edwards model, modulo the nonuniversal rescaling of abscissa and ordinate:

$$f_R(x) = K_1 \alpha_R^2(K_2 x) \qquad (11)$$
$$f_S(x) = (K_1/6) \alpha_S^2(K_2 x) \qquad (12)$$
$$f_A(x) = (K_1^{d/2}/2)(d/2\pi)^{-d/2} \tilde{h}(K_2 x) \qquad (13)$$

Here $\alpha_R^2(z)$, $\alpha_S^2(z)$ and $\tilde{h}(z) \equiv zh(z)$ are the conventional expansion and second virial factors of the continuum Edwards model [2, 5, 6, 13, 14], and $K_1$ and $K_2$ are nonuniversal scale factors. Thus, the modern two-parameter theory (continuum Edwards model) *is* a correct theory for a certain limiting regime in the molecular-weight/temperature plane — but this regime is *not* the one previously thought [24].

The reasonably good agreement that has been observed [25] between the modern two-parameter theory and *some* (but not all [18]) experiments on dilute polymer solutions is due to the fact that *these* experiments (but not the others!) were performed rather near the theta temperature [26]. Unfortunately, the distinction between the ways I and II of taking the long-chain limit has not been clearly understood in the past, and as a result the available experiments (most of which are 15–25 years old) are rather unsystematic in their coverage of the molecular-weight/temperature plane [27]. It would be very interesting to redo the experiments, using modern light-scattering instrumentation [28] and covering systematically the molecular-weight/temperature plane in order to disentangle the limiting regimes I and II.

Another consequence of (11)–(13) is that the temperatures $T_{\theta,\text{eff}}(N)$ defined by $A_2 = 0$ or by $\langle S^2 \rangle/\langle R^2 \rangle = 1/6$ are *not* shifted from $T_\theta$ by a term of order $N^{-\phi}$ (contrary to some previous predictions [29]): the special point in the Edwards model lies at $z = 0$, hence $x = 0$. It is not clear to me whether the exponent $\psi \ (> \phi)$ governing this shift is universal [30].

Let me now explain why (11)–(13) should be true. The correspondence between crossover scaling functions and continuum field theories is best understood in a Wilson-type renormalization-group framework. Consider a Wilson-type RG map $\mathcal{R}$ acting on the (infinite-dimensional) space of Hamiltonians for a field theory or polymer model with some fixed ultraviolet cutoff $\Lambda$ (e.g. on the lattice). Assume that there exists a critical fixed point $H^*$ with stable manifold $\mathcal{M}_s$ and unstable manifold $\mathcal{M}_u$. (For simplicity assume that there are no marginal operators.) Continuum limits are obtained by taking a sequence of initial Hamiltonians $H_n$ approaching the stable manifold, and rescaling lengths by suitable factors. This rescaling is equivalent to applying the map $\mathcal{R}$ a suitable number of times. The low-energy effective Hamiltonians $H_n^{eff} \equiv \mathcal{R}^n H_n$ then tend to the *unstable manifold* (Fig. 1). Continuum field theories $\mathcal{F}$ are thus in one-to-one correspondence with Hamiltonians $H$ on the unstable manifold: the correlation functions of $\mathcal{F}$ at momenta $|p| \leq \Lambda$ are *equal* to the correlation functions of $H$ with cutoff $\Lambda$. This point of view has been emphasized by Wilson and Kogut [31] and others [32, 33].

Suppose that the unstable manifold $\mathcal{M}_u$ is $r$-dimensional, i.e. there are $r$ relevant operators with exponents $\lambda_1 \geq \lambda_2 \geq \ldots \geq \lambda_r > 0$. Then there is an $r$-parameter family of continuum field theories. However, this family is mapped into itself by spatial dilations, so there is really only an $(r-1)$-parameter family of *inequivalent* theories. The continuum field theories can thus be parametrized by a mass scale together with the limiting (as $n \to \infty$) values of the ratios $(g_2/g_1^{\lambda_2/\lambda_1}, \ldots, g_r/g_1^{\lambda_r/\lambda_1})$, where $(g_1, \ldots, g_r, \ldots)$ are the coordinates of the initial Hamiltonians $H_n$ with respect to a set of nonlinear scaling fields. In particular, if $\mathcal{C}$ is any $r$-dimensional manifold which cuts the stable manifold transversally [34], then there is a one-to-one correspondence between asymptotic paths of approach to $\mathcal{M}_s$ within $\mathcal{C}$ and continuum field theories. The crossover scaling functions for approach to the critical point from within $\mathcal{C}$ are thus given precisely (with respect to the nonlinear scaling fields) by the continuum field theories. Of course, the nonlinear scaling fields are connected with the usual physical parameters by a smooth transformation; this transformation gives rise to the nonuniversal scale factors as well as to the analytic corrections to scaling [35].

If $H^*$ is a Gaussian fixed point, then the ratios $(g_2/g_1^{\lambda_2/\lambda_1}, \ldots, g_r/g_1^{\lambda_r/\lambda_1})$ are nothing other than the bare coupling constants (in the field-theoretic normalization convention), normalized by appropriate powers of the physical mass, for the corresponding superrenormalizable field theory.

Thus, I claim that the scaling functions for crossover from the Gaussian to the $n$-vector fixed point (in dimension $3 \leq d < 4$ and zero magnetic field) are given precisely (modulo two nonuniversal scale factors) by the correlation functions of the superrenormalizable $n$-component $\varphi^4$ field theory written in terms of the bare $\varphi^4$ coupling constant. Translated into polymer language (in a fixed-length ensemble), this gives (11)–(13). [In $d = 3$ the $\varphi^6$ interaction is marginally irrelevant; this induces a multiplicative logarithmic correction $x = N^{1/2}(\log N)^{3/22}(T - T_\theta)$, but the conclusion is otherwise unchanged.]

Of course, like all treatments of "general" RG theory, the foregoing discussion is based on the *assumption* that there exists a smooth RG map with a suitable fixed-point structure, etc. It is a highly nontrivial problem to devise a *specific* renormalization-group transformation and prove that it has the requisite properties [33, 36]. But it is a reasonable working hypothesis that this can be done, absent evidence to the contrary.

This viewpoint also sheds light on the connection between the Wilson and field-theoretic renormalization groups. The homogeneous field-theoretic RG equations [37] describe how a family of continuum field theories is mapped into itself under spatial dilation. On the other hand, the continuum field theories are in one-to-one correspondence with Hamiltonians on the unstable manifold; and this correspondence takes spatial dilation into the RG map $\mathcal{R}$. Thus, the field-theoretic RG is nothing other than the Wilson RG *restricted to the unstable manifold* and then rewritten in terms of "renormalized" parameters. This interpretation is probably not new [38], but I have not seen it anywhere in print. Of course, it remains to work out the details for specific RG maps $\mathcal{R}$.

Returning to the polymer problem, we see that the scaling functions $f_R(x)$, $f_S(x)$ and $f_A(x)$ are necessarily nonanalytic at $x = 0$ (since the Edwards model has this property at least for $d > 2$). The scaling functions for $x < 0$ are controlled by the crossover to a *different* fixed point, and they are thus given by a *different* manifold of continuum field theories. Unfortunately I am unable to say much about what these theories might be. Roughly speaking one would expect a $\varphi^6 - \varphi^4$ theory in the symmetry-breaking region (or in polymer language an Edwards model with two-body attraction and three-body repulsion); but since the $\varphi^6$ coupling (or three-body interaction) is a dangerous irrelevant variable [39] in dimension $d \geq 3$, it appears that such continuum theories do not exist. (The situation is analogous to $\varphi^4$ theories in the broken phase in dimension $d \geq 4$ [40].) Possibly the scaling functions are then given by a suitable Landau theory. The scaling functions are presumably continuous at $x = 0$, but I see no reason for all of their derivatives to be continuous.

In dimension $d < 3$, the theta point is no longer Gaussian, and the crossover scaling functions are no longer given by the Edwards model. However, one still expects the scaling functions to be given by two continuum field theories — one for $x > 0$ and one for $x < 0$ — with nonanalyticity at $x = 0$. It would be interesting to examine the scaling functions in exactly-soluble models such as the interacting partially-directed self-avoiding walk [41].

The analysis given here should apply also to magnetic and fluid tricritical points [42].

I wish to thank Jim Barrett, Richard Brak, Sergio Caracciolo, Roberto Fernández, Michael Fisher, Tony Guttmann, Ross Hallett, Gérard Jannink, Antti Kupiainen, Bin Li, Neal Madras, Bernie Nickel, Andrea Pelissetto, Stu Whittington and especially Bertrand Duplantier for very helpful conversations and/or correspondence. This research was supported in part by National Science Foundation grant DMS–9200719, Department of Energy contract DE-FG02-90ER40581, NATO Collaborative Research Grant CRG 910251, and a grant from the New York University Research Challenge Fund. Acknowledgment is also made to the donors of the Petroleum Research Fund, administered by the American Chemical Society, for partial support of this research.

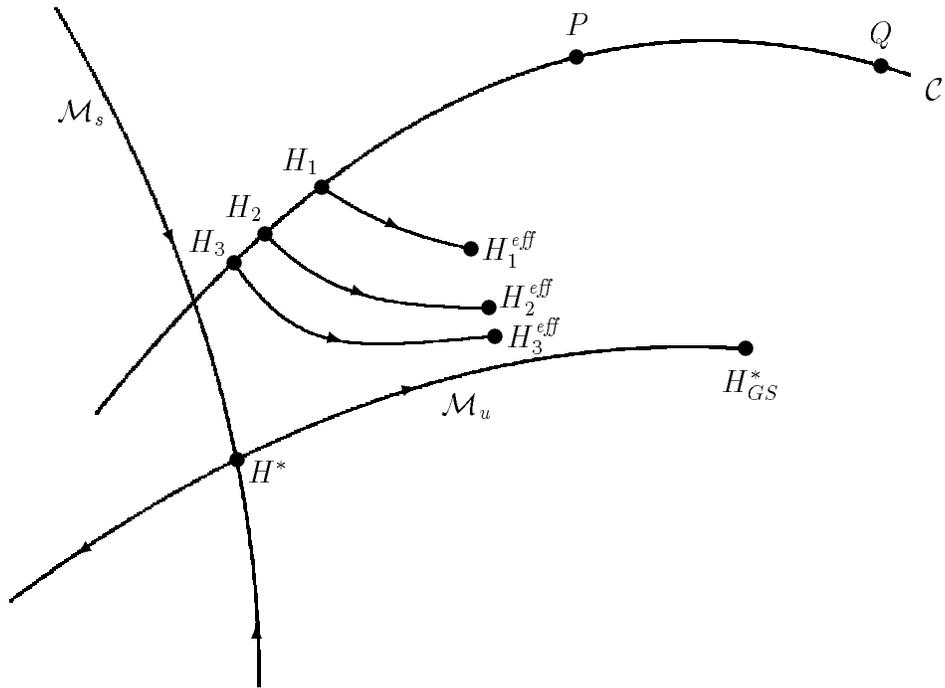

Figure 1: Renormalization-group flow in the neighborhood of a fixed point $H^*$. $\mathcal{M}_s$ (resp. $\mathcal{M}_u$) is the stable (resp. unstable) manifold. Case Ia: Models in the good-solvent regime may have correction-to-scaling amplitudes that are either negative ($P$) or positive ($Q$). Case II: The initial Hamiltonians $H_n$ approach the stable manifold, while the low-energy effective Hamiltonians $H_n^{eff} \equiv \mathcal{R}^n H_n$ approach the unstable manifold.